\documentclass[preprint2]{aastex63}

\usepackage{amsmath}
\begin{document}

\title{Zwicky Transient Facility Observations of Trojan Asteroids: A Thousand Colors, Rotation Amplitudes, and Phase Functions}
\correspondingauthor{Michael Brown}
\email{mbrown@caltech.edu}

\author{Madeline Schemel}
\affiliation{Division of Geological and Planetary Sciences\\
California Institute of Technology\\
Pasadena, CA 9125, USA}

\author[0000-0002-8255-0545]{Michael E. Brown}
\affiliation{Division of Geological and Planetary Sciences\\
California Institute of Technology\\
Pasadena, CA 9125, USA}

\begin{abstract}
We introduce a new method for analyzing sparse photometric data of asteroids and apply it to Zwicky Transient Facility observations
of the Jupiter Trojan asteroids. The method relies on the creation
of a likelihood model that includes the probability 
distribution of rotational brightness variations
at an unknown rotation phase. The likelihood model
is analyzed via a Markov Chain Monte Carlo to quantify 
the uncertainty in our parameter estimates. Using this 
method, we provide color, phase parameter, 
absolute magnitude and amplitude of rotation measurements 
for 1049 Jupiter Trojans. We find that phase parameter is correlated with color and 
the distribution of Trojan asteroid rotational amplitudes is indistinguishable from that of main-belt asteroids. 

\end{abstract}

\section{Introduction}
The Jupiter Trojan asteroids, minor bodies that share Jupiter's orbit, are of interest due to important questions surrounding their origin. More specifically, recent models of the early solar system challenge the traditional idea of smooth planetary migration and instead raise the idea of dynamical instability, where the orbits of planets and minor bodies migrated
in the early solar system 
\citep{nice}. In these models, the Jupiter Trojans were formed not at their current distance, but instead are objects
captured from the source population of the 
Kuiper Belt objects \citep{morbidelli, roig_migration}.

Limited data measuring important physical parameters of the Jupiter Trojans exist. Photometric measurements have revealed a bimodality in the
color distribution of these objects, and are generally used to separate 
the Trojans into two distinct populations, termed the 'red' and 'less-red' groups \citep{emery_2011, wong_2016}.
Other measurements, such as light curve amplitude and phase have only been performed on small numbers of Jupiter Trojans. 
\citet{Shevchenko_phase} and \citet{Schaefer_phase} measure the phase curves of three and nine Jupiter Trojans, respectively. \citet{Mottola_2011}, \citet{French_2015}, \citet{Ryan_2017} and \citet{Szabo_2017} all report amplitude of rotation and period results from rotational lightcurves studies. These papers report results on between 19 and 80 Jupiter Trojans. 

The recent abundance of large astronomical surveys has provided astronomers with the unique opportunity to measure properties of large numbers of minor solar system bodies in bulk. Such bulk measurements allow extraction of large scale properties, identification of extreme outliers, and comparisons between different populations of minor solar system bodies. 
For example, analysis of colors of 10,592 asteroids from the Sloan Digital Sky Survey (SDSS) has  been used to determine the colors of asteroid families \citep{Ivezic_2002}. Similarly, analysis of ~ 250,000 asteroids from the PanSTARRS catalog has shown a relationship between phase parameters and asteroid taxonomy \citep{panstarrs_fits}. 
\citet{ptf_lcs} obtained rotational period fits to 8,300 asteroids from the Palomar Transient Facility catalog. 
 They were able to compare amplitude and period distributions for different asteroid types and identify a number of light curves that displayed 
cusp like minima suggestive of binary system.

While such surveys are a potential rich source of information, the cadence of data collection from these surveys often makes extracting quantitative information about solar system bodies difficult. Most small bodies in the solar system are irregularly shaped and thus vary in observed brightness as they rotate and reflect higher or lower cross-sectional areas. Unfortunately, neither the shapes nor the rotation periods of most bodies is known, so the measured brightnesses occur at unknown rotational phases, making comparison of measurements to derive colors, rotation amplitudes, or phase functions difficult. This problem is generally referred to as the problem of sparse photometry, and potential solutions have been discussed for at least a decade (Warner and Harris 2011; Hanuš and Ďurech 2012; Waszczak et al. 2015), but all suffer from one serious pitfall: one must be able to measure the rotation period of the object to be able to use the photometry. \citet{panstarrs_fits} attempt
to solve this problem by using a 
Monte Carlo method on Pan-STARRS data to sample many potential
rotation periods and amplitudes and derive absolute magnitudes.
While promising, the method depends on asteroid population
models and produces large uncertainties.

Here, we develop a method of extracting quantitative 
parameters from sparse photometry that does not rely on 
being able to measure the rotational period and makes no
assumptions about the underlying population statistics. 
The key insight is that, regardless of the 
rotational period, the probability distribution of the brightness
of an object at a random rotational phase is a quantifiable function of only the asteroid shape.

We use this new method
to analyze photometry of Jupiter Trojan asteroids obtained from the publicly released data of the Zwicky Transient Facility (ZTF). We extract absolute magnitudes, colors, solar phase function parameters, and rotational amplitudes for 1054 such objects. 
 In Section \ref{data}, we discuss ZTF data in more detail. We discuss our model for Trojan brightness in Section \ref{likelihood model} and, in Section \ref{data analysis}, we detail the Markov Chain Monte Carlo (MCMC) method used to extract important parameters. Section \ref{results} reports the results of our analysis and discuss overall population trends.

\section{Data}\label{data}
The Zwicky Transient Facility (ZTF) is a time-domain survey run on the 
Samuel Oschin 48” Schmidt telescope at Palomar Observatory. The CCD 
camera covers a 47 deg$^2$ field of view to reported limiting magnitude of $g=20.8$ and $r
=20.6$ in the typical 15 second exposures \citep{ztf_overview}. The public survey 
attempts to cover the observable northern sky 
every three nights in both the $g$ and $r$ bands.
The same region of sky is rarely covered more than
once in a single night. Every morning data from the
previous night are compared to template images and all transient or variable 
objects are reported at ztf.uw.edu/alerts/public.
A typical night 
will report tens of thousands of
transient detections, many of which are moving 
objects in the solar system.

Along with photometric parameters of each transient detection (including apparent magnitude, magnitude uncertainty, and a 5$\sigma$ limiting magnitude 
threshold), the ZTF pipeline
reports the closest known solar system object at the moment of observation, 
the angular distance to this
object, and the predicted brightness of this object. To extract a database
of all Trojan asteroid detections, we search for all detections within 5
arcseconds of numbered Jupiter Trojans (we restrict ourselves to numbered 
objects to guarantee that the orbits are known well enough that the predicted position is accurate). From 
1 June 2018 until 13 Aug 2020, we find a total of 59961 detections of 2743 numbered Trojan asteroids. We limit our analysis to the 1054 Trojan asteroids with greater than 10 detections.

\section{Likelihood model}\label{likelihood model}
The apparent magnitude of a Trojan asteroid in a particular filter band is a
function of its unknown absolute magnitude, color, solar phase function,
and the amplitude and phase of its rotation
light curve, and its known distance from the sun and earth.
Given a set of parameters, we can predict the brightness 
of an object at the time of observation. 
If we exclude rotation for the moment, the predicted magnitude, $m_p$,
can be calculated as:

\begin{equation}
    m_p =H_r + c - 2.5log_{10}[\phi(\alpha)]+5\log_{10}[R\Delta], \\
\end{equation}
where $H_r$ is the absolute magnitude
in the $r$ band, $c$ is the color term, which is zero for observations in the $r$-band and the $g$-$r$ color for observations in the $g$-band, 
$\phi(\alpha)$ is the solar phase function with $\alpha$ 
the solar phase angle, 
and $R$ and $\Delta$ are the heliocentric and geocentric 
distances, respectively. 
We use the H-G phase function proposed by \citet{Bowell_phase} and adopted by the IAU:
\begin{equation}
    \phi(\alpha) = (1-G)\phi_{1}(\alpha) + G\phi_{2}(\alpha),
\end{equation}

where 
\begin{equation}
\phi_{i} = \exp[-A_i \tan^{B_i} \frac{1}{2} \alpha]
\end{equation}

and $A_1 = 3.33$, $A_2 = 1.87$, $B_1 = 0.63$ and $B_2 = 1.22$ \citep{muinonen_2010}.

This one parameter phase function provides unique fits to our limited data and we found it to be more accurate to our data than a linear fit. It has also been widely cited in literature, which allows us to easily compare our results with previous studies. 

The (unnormalized) probability of making an observation of apparent magnitude $m$ given a true magnitude $m_p$ can be calculated as: 
\begin{equation}
    P_g(m| m_p, \sigma) = \exp[\frac{-(m_p-m)^2}{2\sigma^2}],
\end{equation}
where $\sigma$ is the uncertainty in the observation and is assumed to
be a Gaussian.  
Adding the effects of rotation to 
our model
traditionally requires knowing the rotational period so that 
the phase at the 
time of the observation can be known. We circumvent this
difficulty by avoiding fitting for a rotational phase,
but, instead, 
take into account 
the probability of making an observation, $m$, given a  known rotational amplitude but an arbitrary phase. 

We determine the probability distribution function of
rotational brightness by assuming that the light curve is
a sinusoid,
which appears adequate for most Trojan light curves \citep{Szabo_2017, Ryan_2017}
but will fit less well some extreme light curves or those with two
peaks of differing rotational amplitudes. 
The effects of these limitations are
discussed below. 

We assume the light curve is a sine function of the form

\begin{equation}
    m = \frac{A}{2}sin(p) + m_p,
\end{equation}
where 
$A$ is the peak-to-peak variation in the light curve, and
$p$ is the (unknown) rotational phase. One can easily show that the
(unnormalized) probability distribution function of 
$m$
over a full rotational
period is simply
\begin{equation}
P_r(m|m_p,A)=\frac{1}{(A/2)^2-(m - m_p)^2}.
\end{equation}

A key assumption here is that the amplitude of the light curve
does not change over the time period of analysis. While this assumption
is adequate for distant objects such as Trojans whose viewing
geometry does not change quickly, such an assumption is clearly 
unsatisfactory for analysis of closer populations such as main-belt
asteroids. 

Next, we convolve the rotational probability distribution function with 
the Gaussian distribution for the uncertainty:
\begin{equation}
P(m|m_p, A, \sigma)=P_r(m|m_p,A)\circledast P_g(m|m_p, \sigma),
\end{equation}
where $P$ is the full probability distribution function including
rotation and uncertainty and $P_r$ and $P_g$ account for the
rotation and the Gaussian uncertainty, respectively.

\begin{figure}
\plotone{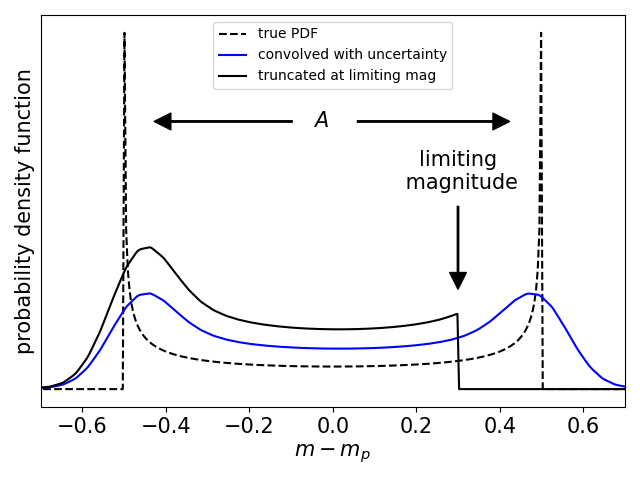}
\caption{An example shape of the probability distribution
function for a single data point. For this example,
the mean magnitude is centered at 0, with a peak-to-peak
rotational amplitude of 1.0 mag. The dashed line represents
the rotational probability distribution function,
and the solid blue line shows the distribution once it has been 
convolved with a gaussian for uncertainty. 
The solid black line shows the effects of an observation 
near the limiting magnitude, which we assume here to be 0.3
magnitudes fainter than the mean magnitude detection.}
\label{fig: fig1}

\end{figure}

One further difficulty is that ZTF observations only report positive detections. We do not 
get upper limits for times of non-detections. Asteroids observed 
close to the detection limit of ZTF could have consistently underestimated rotational
amplitudes if they are only seen at the brightest portion
of their phase curve. To account for this difficulty, 
take the ZTF 5 $\sigma$ limiting magnitude
threshold reported for each detection as the magnitude
below which the detection would not have been reported.
We then truncate the probability distribution 
function at the
limiting magnitude of each observation and renormalize the
area of the PDF to unity.  
The practical effect
of this truncation will be that observations close to the magnitude
limit will provide little constraint on the amplitude of
rotation, likely leading to upper limits for the modeled 
rotational amplitude and correctly reflecting our true uncertainty in the 
amplitude of the rotation.
The general shape of the resulting distribution can be seen in Figure \ref{fig: fig1}.
This distribution is used to calculate the likelihood of each model parameter as a function of the data. 

Because we are using large automatically generated catalogs with
no manual inspection of the detections, a chance
exists that some of the reported detections will be spurious. 
Some of these spurious detections are easily recognizable and can be removed manually. To do this, we remove all points further than 1.5 mag away from a predicted value prior to running the MCMC. For this initial cut, the predicted value is calculated using the absolute magnitude from the JPL Small-body Dataset\footnote{https://ssd.jpl.nasa.gov/sbdb.cgi}, an assumed phase parameter, G, of 0.15, and the mean $g-r$ color of the population. This removes clear outliers, but spurious detections closer to the
predicted apparent magnitude can still be present. We include
a final parameter, $f$, which is the
probability (per magnitude range observed) 
that an observation is spurious. 

We incorporate this parameter into our model by adding $f* \delta m$ across the entire range of the distribution. Here, $\delta m$ is the small magnitude width  by which the likelihood function is binned. 
As a result, the original distribution must be re-normalized such that the total probability sums to 1. 
The likelihood contribution introduced by accounting for outliers  sums to $f*\Delta m$, where $\Delta m$ is the total range covered by the distribution. 
We have already limited the range $\Delta m$ with our removal of obvious outliers more than $\pm$1.5 magnitudes
from the predicted brightness, so the maximum value
of $\Delta m$ can be 3.0. The value of $\Delta m$ can
be below 3.0 is the limiting magnitude is closer than 1.5
magnitudes to the predicted magnitude.

This results in a final equation of the form:
\begin{equation}
    L(m_p, A, \sigma | m)= (1-f\Delta m)P(m | m_p, A, \sigma) + f 
\end{equation}

For small values of $f$ the likelihood function will still prefer
to fit the data into the higher likelihood region of the rotational
model, but the existence of a true outlier will not generate 
an unrealistically low likelihood.

The viewing geometry of Jupiter Trojans changes by about
30 degrees per year, so while our assumption about the rotational
amplitude being invariant is reasonable for a single opposition,
it begins to break down with time.
We thus separate the observations by opposition season, 
and calculate a distinct amplitude of rotation, $A$,
and a distinct absolute magnitude, for each. 
For absolute magnitude, we set a reference absolute magnitude, $H_r$, as the absolute magnitude of the second opposition (the year 2019), 
as this year generally contains the largest number of observations, and tabulate parameters to measure magnitude offset from the reference for any other oppositions.  
The ZTF survey has been collecting data for more than two years, 
meaning that, for most Trojan asteroids, we measure at least two and 
sometimes three separate rotational amplitudes. 

Our full likelihood model thus has as many as 9 parameters, 
including the 2019 average absolute magnitude ($H_r$), offsets 
to the 2019 magnitude for the 2018 and 2020 average magnitudes 
(off$_1$ and off$_3$), rotational amplitudes for 2018, 2019,
and 2020 ($A_1$, $A_2$, and $A_3$), a single color $g-r$, a 
single phase parameter $G$, and a fraction of outliers per 
magnitude, $f$.

\section{Data analysis}\label{data analysis}
With the existence of a likelihood model describing the data we
can now attempt to determine the parameters which fit the data for
each individual Trojan asteroid. While a simple 
Maximum Likelihood approach would yield optimal parameters for
each object, we instead use a Markov Chain Monte Carlo (MCMC)
model to allow us to appropriately incorporate sensible priors and
to better understand uncertainties via the posterior distributions. 

Specifically, we use the affine invariant method of \citet{emcee}  implemented in Python using the {\it emcee} package \citep{fm2012emcee}.
This MCMC method is an
ensemble sampler which uses 
a set of n {\it walkers} to sample the distribution of
parameter space.
. 
For this analysis, the ensemble sampler comprised of 30 individual walkers. 
% typical autocorrelation time around 100
Examination of the trace plots shows that the chains generally converge within the first 500 steps. 
Typical auto-correlation times are ~100, but can vary depending on the asteroid and parameter. 
For our final analysis each walker completed 5,000 steps, the first 1,000 of which were discarded. 
This results in a total of 120,000 points sampling each posterior distribution. 
 We found this to be a reasonable number of samples to ensure convergence of the MCMC while limiting the amount of 
 computation time required. 

We incorporate priors on all
of our parameters. For the
phase parameter, $G$, the $g-r$
color, $gr$, and the rotation amplitude $A$, we assume uniform
priors over the following ranges:

\begin{equation}
\begin{split}
-0.25 < G < 1.0 \\
-1.5 < gr < 1.5\\
0 < A < 2\\
\end{split}
\end{equation}

All of these Jupiter Trojans have 
previously measured absolute magnitudes, so we seek a prior
on absolute magnitude reflecting this (imperfect)  knowledge.  
A preliminary run on the ZTF data with only a uniform prior on $H_r$ indicated the median offset between our measured r-band absolute magnitudes and the absolute magnitudes (in the $V$ band) tabulated
by the JPL Small-body Dataset was 0.04
magnitudes. 
For the our final run, we then set the prior on $H_r$ to be a gaussian with a mean at $M_{JPL} + 0.04$ and a standard deviation of 0.5, reflecting the spread observed in
our initial run. 
The prior on off$_1$ and off$_3$, the change in absolute magnitude from one opposition to the next, is a gaussian with a mean at 0 and a standard deviation of 0.1. We expect the magnitude offset between oppositions to be small, given that the light curve must change continuously from season to season as the viewing geometry of asteroid changes. Empirically, we find the offset is generally less than 0.2 in either direction. The specific values of all of the
priors have little effect on the final results except
in the cases where not enough data points exist to
constrain parameters, in which case the corresponding
uncertainties reflect out poor knowledge of the 
derived parameters.

For $f$ we strongly bias the prior to a small number of outliers
by using the function $
  p(f)  = cos(f\pi/2)^{10}$
 We also truncate $f$ in order to ensure that the likelihood does not sum to less than zero. Because the maximum range of $\Delta m$ is 3 mag (due to our cutoff of large outliers) we limit $f$ to values below 0.33.

\begin{figure}
\plotone{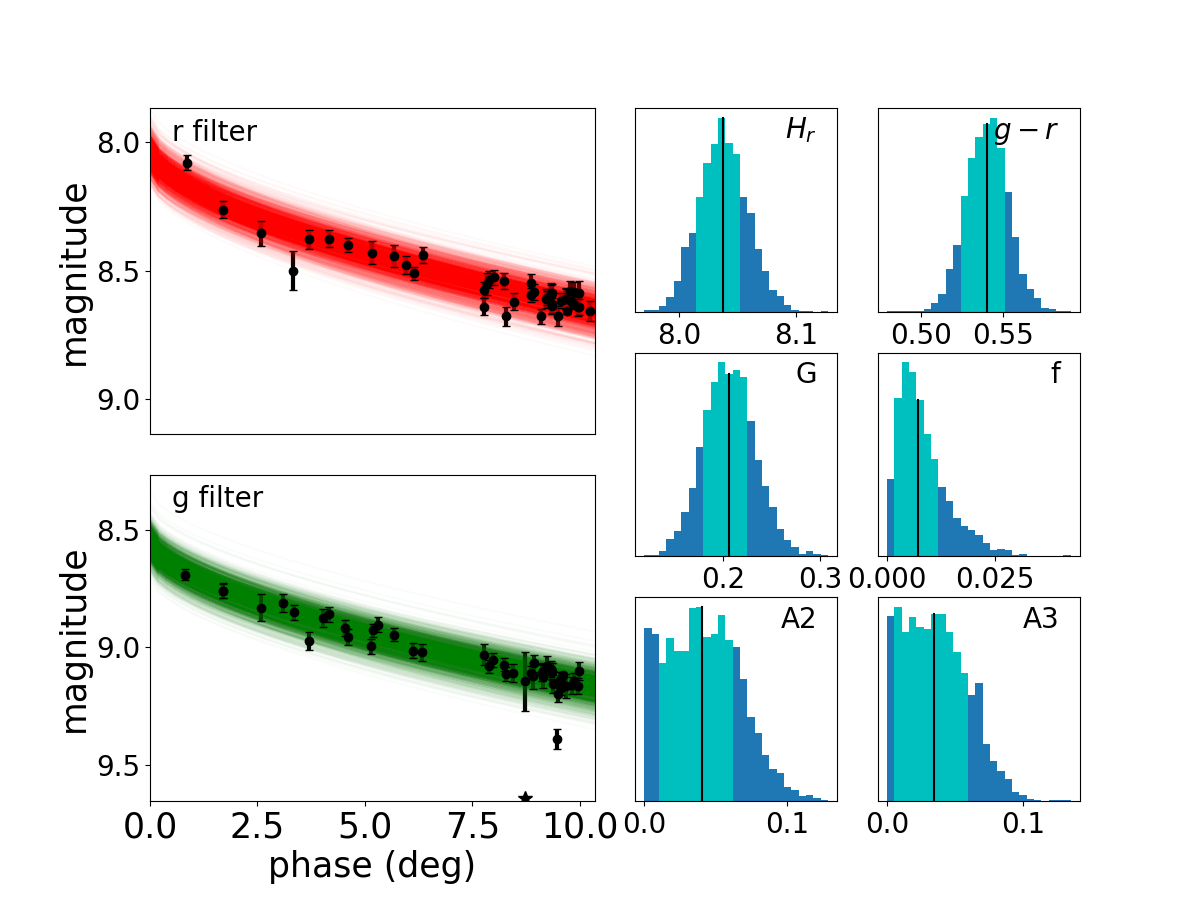}
\caption{The model fit to ZTF observations of (617) Patroclus. The data and uncertainties for the $g$ and $r$ filters are plotted in green and red, respectively. The detection limit of each observation is denoted by the star symbol. The phase curves shown randomly selected samples from the MCMC and thus
approximately represent the probability distribution of 
expected magnitudes. When the results are well constrained, the density of curves will match the density of data points at each magnitude. 
The resulting posterior distribution of each parameter is shown to the right of the phase curve fits. The median value is indicated with the black arrow. The range of light blue colors represents covers the 68\% of data with the
highest density.}
\label{fig: plot_617}
\end{figure}

\begin{figure}
\plotone{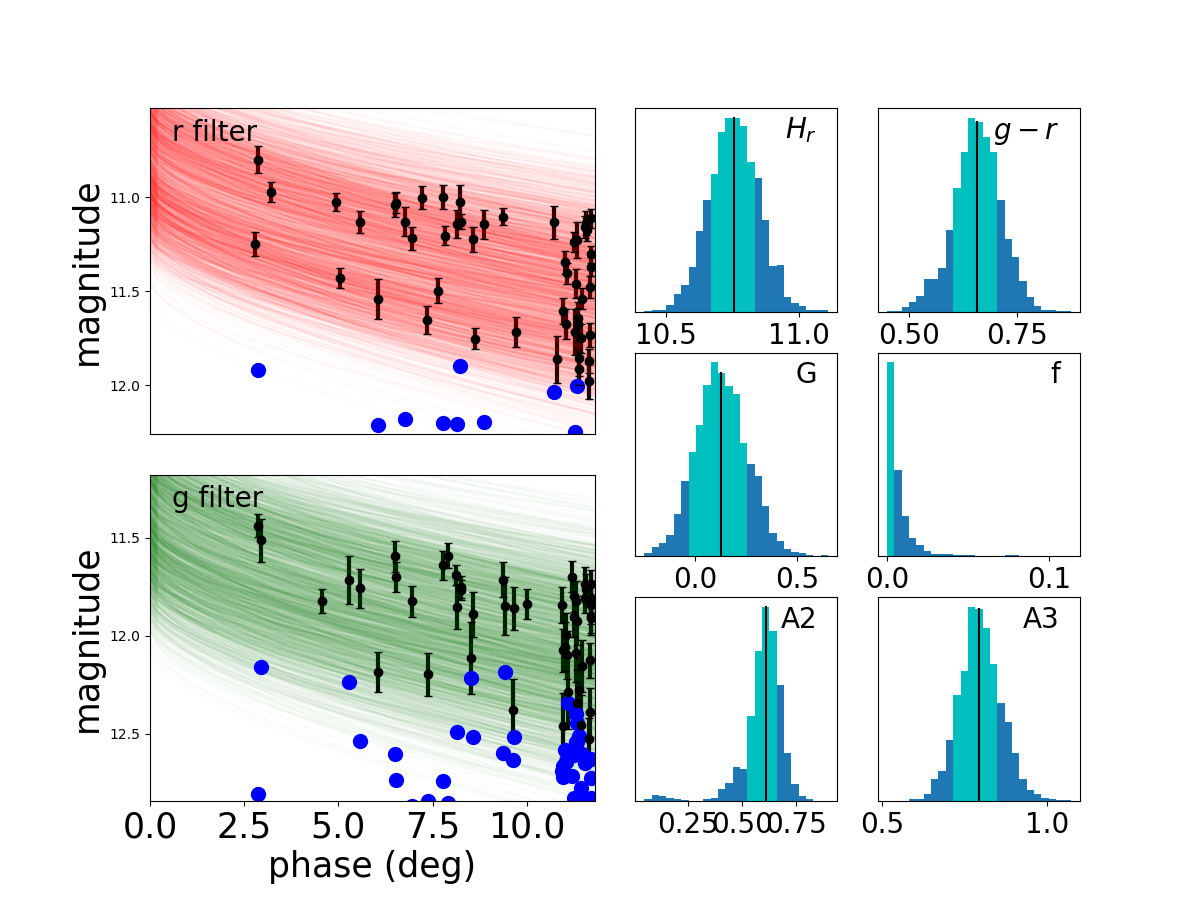}
\caption{The model fit to ZTF observations of (11351) Leucus.
The data, models, and results are as in Figure 2.}
\label{fig: plot_11351}
\end{figure}

\begin{figure}
\plotone{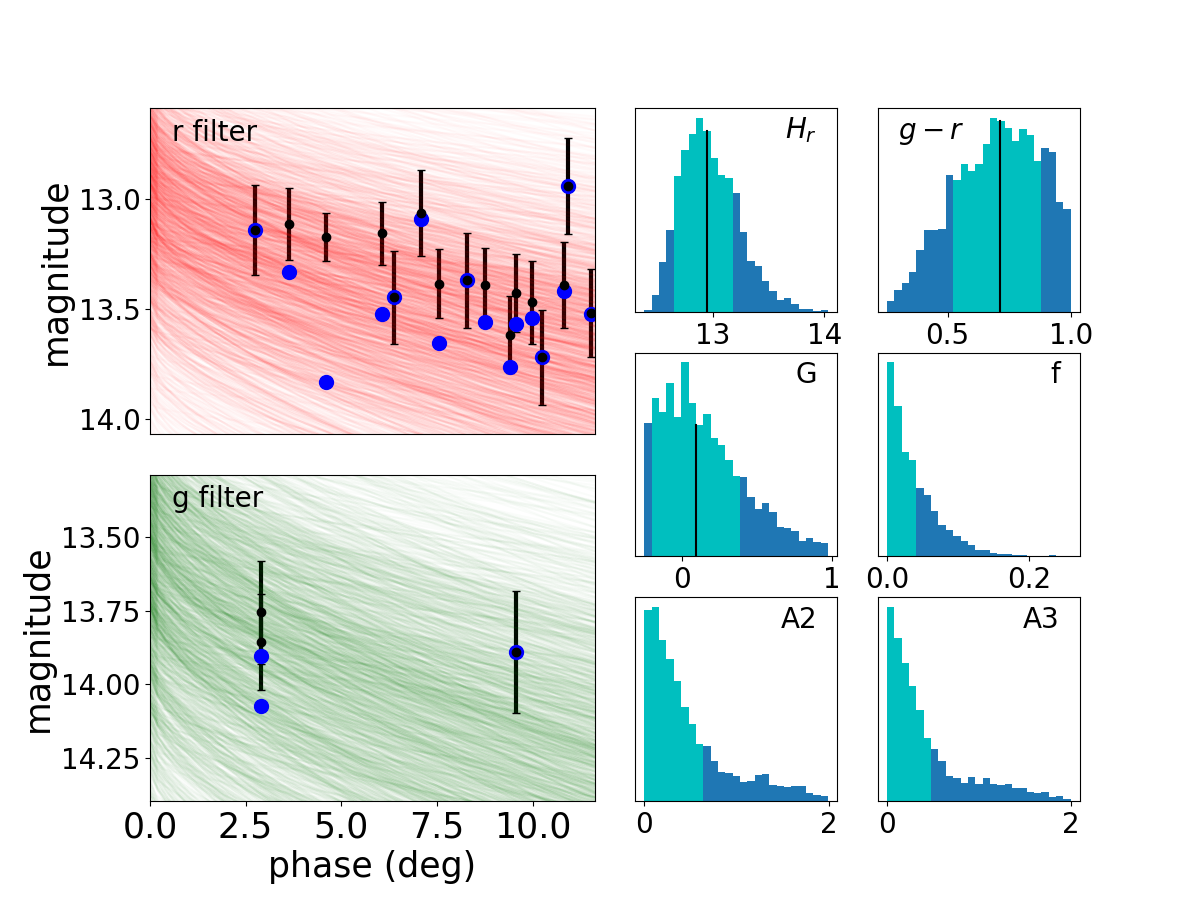}
\caption{The model fit to ZTF observations of Trojan asteroid 421382.The data, models, and results are as in Figure 2.}
\label{fig: plot_421382}
\end{figure}

We show example fits to Trojans (617) Patroclus, (11351) Leucus and (421382) in Figures \ref{fig: plot_617}, \ref{fig: plot_11351} and \ref{fig: plot_421382}, respectively. 
The measured parameter values for each asteroid are shown in Table 1. 

The large bright Trojan (617) Patroclus was observed 90 times by ZTF (44 in the r filter and 46 in the g filter), which is on the higher end for the Trojan asteroids examined in this study, giving tightly constrained results (Figure 1). 
It was observed in three oppositions and, as a result, three rotational amplitude values were measured. The rotational amplitudes for the 2019 and 2020 oppositions are shown in Figure \ref{fig: plot_617}. Two offset values, those between the 2018 and 2019 oppositions and between the 2019 and 2020 oppositions, were measured. The rotational amplitude and mean absolute magnitude changes
little from season to season. 

(11351) Leucus is an example of a Jupiter Trojan with a larger rotational amplitude (Figure 2). The bifurcation between the points clustered at the top and bottom of the lightcurve is clearly visible. This asteroid was also observed in three oppositions.

(421382) 2013 UE4 is an example of the results for a less well-constrained fit (Figure 3). 
There are a limited number of detections, most of which are close to the observing limit, so the analysis 
correctly captures the possibility that the rotation
amplitude could be higher but with many non-detections.
As a result, unlike in the previous two cases, the MCMC samples are
not symmetric about the data, but rather reflect the possibility of undetected points below the observation limit.

\begin{deluxetable}{cccc}[h!]\label{table: Table 1}
\tablehead{\colhead{} & \colhead{(617) Patroclus} & \colhead{(11351) Leucus} & \colhead{(421382) 2013 UE4}}
\startdata$A_1$ & $0.07 \pm 0.03$ & $0.7 \pm 0.1$ & $< 1.2$\\
$A_2$ & $0.04 \pm 0.03$ & $0.6 \pm 0.1$ & $< 0.7$ \\
$A_3$ & $0.03 \pm 0.01$ & $0.8 \pm 0.1$ &  $< 0.6$\\
G & $0.20 \pm 0.03$ & $0.1 \pm 0.1$ & $0.1 \pm 0.3$\\ 
$H_r$ & $8.04 \pm 0.02$ & $10.8 \pm 0.1$ & $12.9\pm  0.3$\\
off$_1$ & $0.00 \pm 0.02$ & $-0.01 \pm 0.05$ & $0.0 \pm 0.1$\\
off$_3$ & $0.08 \pm 0.01$ & $0.02 \pm 0.05$ & $0.05 \pm 0.09$\\
g-r & $0.51 \pm 0.01$ & $0.58 \pm 0.04$ & $0.6 \pm 0.1$\\
f & $0.0007 \pm 0.0005$ & $<0.004$ & $< 0.02$\\
\enddata
\caption{Results for Trojan asteroids (617) Patroclus, (11351) Leucus, and  2013 UE4 (421382).}
\end{deluxetable}

\section{Results and Discussion}\label{results}

The MCMC analysis was run on 1054 Jupiter Trojans, all with ten or more observations in ZTF. Of these, all have measurements of phase parameter 
% need to check this after re-run, might've changed
magnitude, rotational amplitude for at least one opposition, and $g-r$ color. For 1033 of the Trojans we were able to measure the rotational amplitude in at least two oppositions and for 396 we measured the rotational amplitude in all three years in which ZTF has been conducting observations. 

The mean 
and standard deviation of each parameter across all analyzed Trojans is shown in Table 2.

\begin{deluxetable}{c c}[h!]\label{table: Table 2}
\tablehead{\colhead{ Parameter} &\colhead{ Value}}
\startdata
 G & 0.24 $\pm$ 0.16  \\
 $g-r$ & 0.56 $\pm$ 0.06\\
A1 & 0.46 $\pm$ 0.30 \\
A2 & 0.28 $\pm$ 0.27 \\
A3 & 0.30 $\pm$ 0.25 \\
f & 0.02 $\pm$ 0.02 \\
\enddata
\caption{Overall population statistics of parameters}
\end{deluxetable}

The mean values of the phase parameter, $G$ and $g -r$ color are both within expected ranges \citep{emery_2011, phase_taxonomy}. 
These parameter values will all be discussed in more detail later in this section. 
Lastly, the fraction of outliers in the ZTF data is very small. 

We have also included a table of our results containing parameter measurements for all 1054 asteroids, which is available online. A sample of these results is shown in Table 3.

\begin{deluxetable*}{cccccccccc}

\tablehead{
\colhead{Trojan} & \colhead{$H_r$} & \colhead{$G$} & \colhead{$g-r$} &\colhead{ off$_1$\ \ (2018)} & \colhead{ off$_3$\ \ (2020)}& \colhead{  $A_1$\ \  (2018)}&\colhead{  $A_2$\ \ (2019)}&\colhead{$A_3$\ \ (2020)}&\colhead{$f$}
}
\startdata
 588& $  8.09_{-0.01}^{+0.02}$& $  0.35_{-0.03}^{+0.04}$&   0.58$\pm$0.01& - &   0.05$\pm$0.01& - & $<  0.03$& $<  0.02$& $<  0.01$\\
617&   8.04$\pm$0.02& $  0.20_{-0.02}^{+0.03}$&   0.51$\pm$0.01& $  0.00_{-0.02}^{+0.01}$&   0.08$\pm$0.01& $  0.04_{-0.02}^{+0.01}$& $<  0.03$& $<  0.03$& $<  0.01$\\
624&   7.09$\pm$0.03&   0.29$\pm$0.05&   0.61$\pm$0.01& $  0.00_{-0.03}^{+0.01}$& $ -0.16_{-0.03}^{+0.01}$&   0.06$\pm$0.03&   0.08$\pm$0.01& $  0.23_{-0.02}^{+0.01}$& $<  0.01$\\
659&   8.39$\pm$0.05&   0.15$\pm$0.06& $  0.51_{-0.01}^{+0.02}$& - & $  0.05_{-0.03}^{+0.02}$& - & $  0.12_{-0.01}^{+0.02}$& $<  0.05$& $<  0.03$\\
884& $  8.55_{-0.05}^{+0.06}$&   0.39$\pm$0.09&   0.58$\pm$0.03& - & $ -0.12_{-0.05}^{+0.04}$& - &   0.08$\pm$0.02&   0.20$\pm$0.04& $<  0.02$\\
1143&   8.22$\pm$0.02& $  0.33_{-0.05}^{+0.04}$& $  0.58_{-0.01}^{+0.02}$& $  0.00_{-0.09}^{+0.08}$& $ -0.02_{-0.02}^{+0.01}$& $  0.20_{-0.19}^{+0.20}$& $  0.08_{-0.01}^{+0.02}$& $  0.08_{-0.02}^{+0.01}$& $<  0.01$\\
1172& $  7.87_{-0.04}^{+0.05}$&   0.19$\pm$0.06&   0.61$\pm$0.02& - & $ -0.21_{-0.03}^{+0.01}$& - & $  0.12_{-0.01}^{+0.02}$& $<  0.04$& $<  0.02$\\
1173&   8.55$\pm$0.04& $  0.15_{-0.05}^{+0.04}$& $  0.49_{-0.02}^{+0.01}$& - & $  0.07_{-0.01}^{+0.02}$& - & $  0.05_{-0.03}^{+0.02}$&   0.09$\pm$0.01& $<  0.01$\\
1208&   8.77$\pm$0.04&   0.22$\pm$0.05&   0.47$\pm$0.01& $ -0.09_{-0.03}^{+0.01}$&   0.05$\pm$0.02&   0.15$\pm$0.02& $  0.10_{-0.02}^{+0.01}$&   0.06$\pm$0.02& $<  0.01$\\
1404& $  9.14_{-0.05}^{+0.06}$& $  0.22_{-0.09}^{+0.08}$& $  0.48_{-0.03}^{+0.02}$& - & $ -0.05_{-0.03}^{+0.02}$& - &   0.14$\pm$0.02& $  0.10_{-0.02}^{+0.01}$& $<  0.01$\\
1437& $  7.92_{-0.02}^{+0.03}$&   0.17$\pm$0.03&   0.51$\pm$0.01&   0.09$\pm$0.02& $ -0.22_{-0.03}^{+0.01}$& $<  0.03$&   0.05$\pm$0.01&   0.20$\pm$0.02& $<  0.01$\\
1647&  10.36$\pm$0.04&   0.40$\pm$0.07& $  0.60_{-0.01}^{+0.02}$& - & $ -0.02_{-0.03}^{+0.02}$& - &   0.06$\pm$0.02&   0.10$\pm$0.02& $<  0.01$\\
1749&   9.34$\pm$0.02&   0.35$\pm$0.04& $  0.61_{-0.02}^{+0.01}$& $ -0.12_{-0.04}^{+0.03}$&   0.03$\pm$0.03& $<  0.09$& $<  0.02$& $<  0.08$& $<  0.01$\\
1868&   9.41$\pm$0.04& $  0.30_{-0.05}^{+0.06}$&   0.55$\pm$0.01& $  0.09_{-0.02}^{+0.01}$& $ -0.11_{-0.02}^{+0.01}$& $  0.07_{-0.01}^{+0.02}$& $  0.10_{-0.01}^{+0.02}$&   0.16$\pm$0.01& $<  0.00$\\
1869& $ 10.90_{-0.06}^{+0.05}$&   0.20$\pm$0.10&   0.53$\pm$0.03& - &   0.02$\pm$0.03& - & $<  0.04$& $<  0.07$& $<  0.01$\\
1870& $ 10.67_{-0.05}^{+0.06}$& $  0.31_{-0.07}^{+0.08}$&   0.62$\pm$0.03& $  0.02_{-0.10}^{+0.11}$&   0.00$\pm$0.03& $  0.30_{-0.26}^{+0.25}$&   0.17$\pm$0.03& $  0.13_{-0.02}^{+0.03}$& $<  0.01$\\
1871& $ 11.03_{-0.05}^{+0.04}$&   0.18$\pm$0.07& $  0.55_{-0.03}^{+0.02}$& - & $ -0.06_{-0.03}^{+0.01}$& - & $<  0.04$& $<  0.05$& $<  0.01$\\
1872&  10.64$\pm$0.05&   0.09$\pm$0.06&   0.51$\pm$0.02& - &   0.05$\pm$0.03& - & $  0.11_{-0.04}^{+0.03}$& $  0.09_{-0.03}^{+0.02}$& $<  0.01$\\
1873& $ 10.03_{-0.02}^{+0.03}$& $  0.32_{-0.04}^{+0.05}$& $  0.55_{-0.02}^{+0.01}$& - & $ -0.15_{-0.02}^{+0.01}$& - & $<  0.04$& $  0.16_{-0.02}^{+0.01}$& $<  0.01$\\
2146&   9.88$\pm$0.02&   0.21$\pm$0.04&   0.59$\pm$0.02& $ -0.01_{-0.03}^{+0.01}$& $ -0.01_{-0.10}^{+0.08}$& $  0.08_{-0.02}^{+0.01}$& $  0.05_{-0.01}^{+0.02}$& $  0.22_{-0.21}^{+0.20}$& $<  0.01$\\
2148& $ 10.64_{-0.05}^{+0.06}$&   0.12$\pm$0.07& $  0.59_{-0.03}^{+0.02}$& $  0.00_{-0.10}^{+0.09}$& $  0.03_{-0.03}^{+0.02}$& $  0.30_{-0.26}^{+0.25}$& $<  0.06$& $  0.05_{-0.03}^{+0.04}$& $<  0.01$\\
\enddata
\tablecomments{An electronic version of this full table is available online}
\caption{Parameter measurements for all asteroids.}
\end{deluxetable*}

\subsection{Accuracy of Results}
\begin{figure}
\plotone{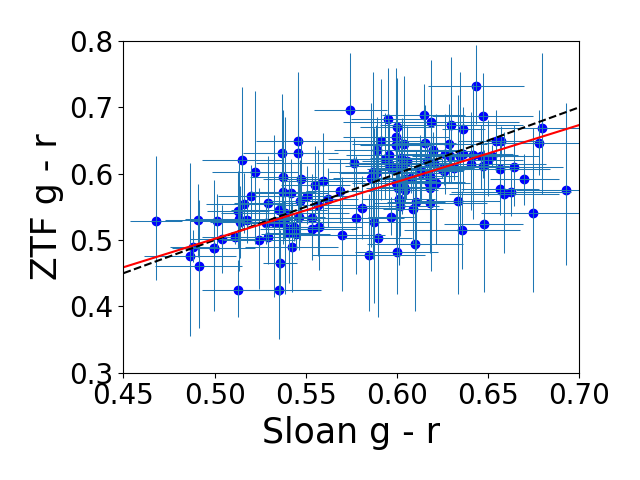}
\caption{A comparison between the $g-r$ colors 
of the 156 Jupiter Trojan
asteroids common to the ZTF
and the Sloan Digital Sky Survey \citet{roig}.The dashed line shows a line of equal
values, while the red line shows that a least square fit
line to the data is nearly identical to the dashed line. }
\label{fig: color_compare}
\end{figure}

To verify our parameter measurements, we compared the results obtained here with other Trojan asteroid surveys in the literature. While there are limited measurements for Trojan phase parameters or light curve amplitudes (partially due to the requirement that direct comparison between rotational amplitude measurements requires both measurements to be from the same opposition period), the color of Trojan asteroids is a fairly well understood property.
We compare our colors measurements to those measured in
the Sloan Digital Sky Survey (SDSS) \citep{roig} for 
the 156 Jupiter Trojans which are common between the
two samples. 
These two data sets show remarkably good agreement, as demonstrated in Figure \ref{fig: color_compare}.
The difference between the two data sets, divided by the
quadrature sum of the uncertainties, is normally distributed around zero, with a sigma of 1.08, compared to a value of
one that would be expected if the two data sets are 
independent measures of the same parameter
with independent normally distributed uncertainties.

% sizing???

Comparison of rotational amplitudes with previous results is 
made difficult by the variation in amplitude from opposition to opposition. Several Trojan asteroids which will be the subject of
a Lucy spacecraft flyby have been closely monitored for the
past few oppositions, however. 
\citet{new_Mottola} measured rotational amplitudes of (11351) Leucus of 0.65 mag for the 2018 opposition and 0.7 mag for the 2019 opposition. We measure rotational amplitudes of 0.73 $\pm$ 0.11 and 0.61 $\pm$ 0.07.
For (3548) Eurybates, 
Mottola (private communication) found a rotational amplitude of 0.1 mag for both oppositions in which we have obtained observations, while we measure upper limits of 0.18 and 0.15 for the 2018 and 2019 oppositions, and he observed a rotational amplitude of (21900) Orus of 0.14 in 2019, when
we find  a rotational amplitude of $0.10 \pm 0.05$.
%For this opposition, the amplitude is 0.15 while we measure an upper limit of   < 0.18. 
Our method appears to produce reliable results both in the case
of well-measured rotational amplitudes and in the upper limits when
no measurement is possible.

\subsection{Phase and Color}

Previous studies of main-belt asteroids have noted differences among asteroid phase parameter between different asteroid taxonomic classes. 
For example, 
\citet{Schaefer_phase} reports differences in the linear phase function 
between different asteroid classes 
%and between different types of minor solar system bodies
while 
\citet{L_M}, \citet{phase_taxonomy} and \citet{panstarrs_fits} all report differences in the Bowell-Lumme phase parameter, $G$, between asteroids of different classes. %the M, S and C classes. 
%Similarly, both \citet{phase_taxonomy} and \citet{panstarrs_fits} report differences in the $G$ phase parameters as a function of asteroid class.

Typically, the Jupiter Trojans are classified as  D and P-type asteroids. 
The average $G$ parameter values for D-type asteroids are $0.23 \pm 0.29$ from \citet{panstarrs_fits} and $0.19 \pm 0.12$ from \citet{phase_taxonomy}. 
Both of these values are within one standard deviation from our mean $G$ value for the Trojan sample. Neither paper quotes an average phase parameter value for P-type asteroids.  \citet{Schaefer_phase}, however, report an average linear slope of 0.039-0.044 for P-type asteroids when looking at phases greater than five degrees and a slope of 0.075-0.113 when looking at phase angles less than 2 degrees. Since most of our observations occur at phases greater than 5, we use those slope values to convert to G parameters. Empirically, linear slopes between 0.039 and 0.044 correspond to G values between 0.34 and 0.42, which are also within one standard deviation from our mean $G$ value for the Trojan sample.

A more appropriate classification scheme for the Trojans, however, is probably to simply use colors. 
%In addition to asteroid type, the Trojan asteroids are commonly classified by color.
We thus examine the relationship between phase parameter and color, looking
to see if the well-known bifurcation of Jupiter Trojans into red and less-red groups is visible and also reflected in the phase parameter.

\begin{figure}
\plotone{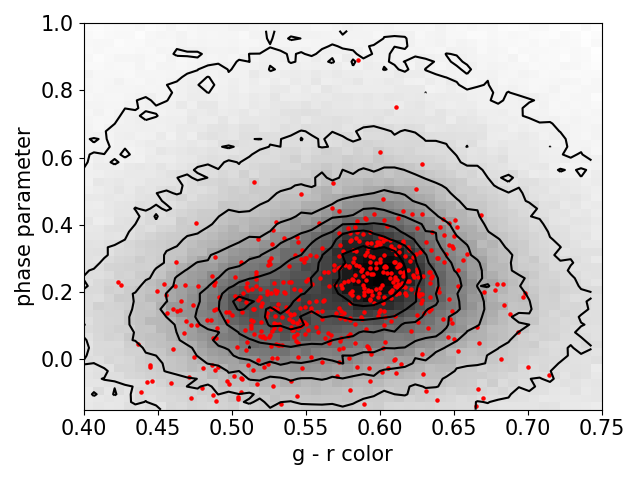}
\caption{We show the relationship between $g - r$ and 
phase parameter, $G$ for the 514 Jupiter Trojans 
with better-than-median uncertainties (red points). 
In the background and contours,
we show the continuous two-dimension
probability distribution function of both of 
these variables for the full data set (see text for 
details).
A positive correlation between the two variables
is apparent.}
\label{fig: G_gr}
\end{figure}

%This relationship is shown in Figure \ref{fig: G_gr}. 
We initially limit our sample to the most well-measured 
Trojans. We do this by selecting the objects where the uncertainty in both G and $g-r$ color is less than the median uncertainty for that parameter. This results in a total of 514 Trojan asteroids. 
These points are shown in Figure \ref{fig: G_gr}, plotted in red. 
For those points, all uncertainties in G are less than 0.27 and all uncertainties in $g-r$ are less than 0.10.

In black, we also plot the G vs $g-r$ relationship for the entire population of analyzed Trojans.
This is done by plotting a distribution of possible G and $g-r$ values.
More specifically, there exists a two dimensional posterior distribution representing the likelihood of a specific pair of G and $g-r$ values.
This is exactly like the one dimensional posterior distributions that show the most likely values for the G and $g-r$ parameters individually, except that it has now been plotted in an additional dimension to show the relationship between the two parameters. 
We can then stack the two-dimensional posterior distributions for each individual Trojan to create a distribution representing the relationship between G and $g-r$ values across the entire population of analyzed Trojans. 
The darker values indicate the values of G and $g-r$ parameters that are more common in our measurements of the Trojan population. 
Contour lines are added to make the delineations more visible.

While uncertainties in our color measurements are large enough that the bimodal distribution of the two color groups is 
only marginally visible, a
correlation between color and phase parameter appears to be present. 
We take the subset of the sample that has uncertainties in color and phase
parameter less than the median values (the points plotted in red in Figure \ref{fig: G_gr}) and perform A Spearman rank correlation test.
This results in a rank correlation coefficient of, $\rho = 0.34$ and a p-value of less than $10^{-5}$, corresponding to a 4.2$\sigma$
correlation between color and phase parameter in the Trojan population.

%It is difficult to pinpoint an specific physical reason for the variation of phase parameter with color. 
%On some level, it is expected that color and phase should be connected. 
%Material composition-- which affects asteroid color -- will also affect the phase parameter \citep{Hapke_1986}. 
%In addition, properties such as grain size, porosity, and large scale surface roughness can affect the phase parameter \citep{lab_phase_curves, muinonen_2002}. 

We suspect that, rather than a linear correlation between color and phase
parameter, we are likely seeing a difference in phase parameter between 
the red and less-red Jupiter Trojans. 
The average $g-r$ color values of the red and less-red Trojan color populations are 0.52 and 0.62 \citep{mag_color_bimodal}. 

We can fit a line to all points with uncertainties less than the median value and solve for the average phase parameter for each color populations.
This results in average phase parameter values of $0.14 \pm 0.05$  and $0.22 \pm 0.07$ for the less-red and red populations. 
The uncertainties in phase parameter were calculated by propagating the error in the slope of the fitted line.

\subsection{Amplitude of Rotation and Diameter}

We compare the amplitude of the rotational 
lightcurve to the diameter.
We use the second opposition (year of 2019), 
because these are generally better constrained. 
In order to appropriately account for objects for which
we have only upper limits and to incorporate the uncertainties
correctly, for each asteroid we add the probability 
density function of amplitude to a bin at the asteroids
diameter. Each diameter bin thus contains the full 
probability density function of amplitude. We then normalize
each diameter bin to allow better comparison
across diameters and present the results
as a two-dimension density function (Figure 7).
No systematic change with
asteroid size is detected.

%We compare to two types of studies.
%In some cases, our comparison is complicated by the 
%in some light curve studies, the goal is to extract the rotational period of the asteroid.
%In the first type our comparison is complicated by the fact that, when no period can be found, no light curve -- and thus no amplitude -- is reported. This biases the studies against low-amplitude light curves, where it is more difficult to extract a period. Our study, meanwhile, does not look for periods at all and, as a result, we do report upper limits for the low-amplitude asteroids in our study. 
%This must be ...

\begin{figure}
\plotone{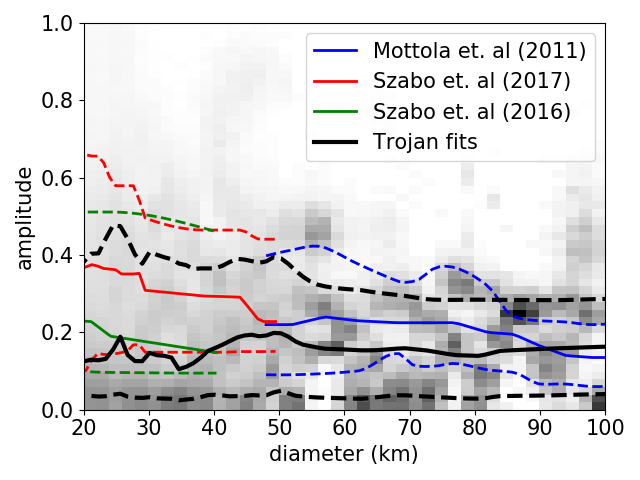}
\caption{Amplitude of rotation distribution of Trojan asteroids
as a function of diameter. The underlying distribution 
shows full probability distribution function of
rotational amplitude in each diameter bin, appropriately
accounting for objects with only upper limits measured. The
solid black line shows the median rotational amplitude
at each diameter, 
while the dashed lines show the 16th and 84th percentiles.
Comparisons to other measurements of Jupiter Trojans
are shown in blue and red, while comparison to main-belt
asteroids is shown in green.}
\label{fig: amp_mag}
\end{figure}

To check our technique, we compared our measurements of rotational amplitude to Trojan asteroid rotational amplitude measurements from \citet{Mottola_2011}, \citet{French_2015}, \citet{Szabo_2017}, and \citet{Ryan_2017}. 
Both \citet{Ryan_2017} and \citet{Szabo_2017} examine the same set of data from the Kepler space telescope, while \citet{Mottola_2011} and \citet{French_2015}  present results obtained using ground based photometry.
The \citet{Mottola_2011} and \citet{Szabo_2017} data are plotted in Figure \ref{fig: amp_mag}, in blue and red respectively.
The results from \citet{Ryan_2017} agree with \citet{Szabo_2017} so are not independently plotted. 
The \citet{French_2015} data (not plotted) covers the same diameter range as the Kepler results, but measures a smaller average amplitude of rotation. 
In general, all of the data are broadly consistent, but the
well-measured Kepler data from \citet{Szabo_2017} and \citet{Ryan_2017} gives mean values approximately 50\% 
larger than ours. We believe that these differences
are due to our assumption of a sinusoidal light curve. 
In the case of a non-sinusoidal light curve, our
technique is more strongly constrained by the smaller
of the two peaks, rather than the full amplitude. 
As a test, we created a non-sinusoidal lightcurve, where the amplitude of the larger peak is 0.25 and the amplitude of the smaller is 0.1. Running our full analysis gives
a rotational amplitude of 0.1. Our amplitudes should thus
be thought of as the smaller of a two-amplitude lightcurve.

We also compare the Jupiter Trojans to the 
larger population.
\citet{ptf_lcs, Szabo_main, polishook} 
%chang, polishook, binzel 
measure light curves and rotational amplitudes for large sets of main-belt asteroids. 
In this study we compare our Trojan rotational amplitudes to those from \citet{Szabo_main}. We use the diameters listed in \citet{lcbd}. Our dataset and the \citet{Szabo_main} dataset overlap in diameter range and, most importantly, the \citet{Szabo_main} paper reports rotational amplitudes even when the period is undetermined. 
Because it is harder to measure an accurate period for asteroids with a low rotational, excluding those asteroids creates a bias against low rotational amplitude asteroids. 
The data from \citet{Szabo_main} are plotted in Figure \ref{fig: amp_mag} in green, and the $\pm$1 $\sigma$ amplitude range overlaps with that from our Trojan fits. 
As such, we report no measurable difference between the rotational amplitude distributions of Trojan asteroids and main-belt asteroids. 

\subsection{Individual objects}
%here is where I would perhaps talk about the high amplitude dudes.
The most potentially interesting outliers in this collection
are those with unusually high rotational amplitudes or long rotation
periods.
A total of 75 objects have 1$\sigma$ lower limits to their rotation amplitude of
0.5 magnitudes or higher for at least one of the three opposition periods.
For each of these objects we subtracted the best fit non-rotating model and
examined the residuals to determine if we could fit an orbital period.
We applied a Lomb-Scargle periodogram analysis to the data and examined
the frequency peak with the highest amplitude.  While many of the data sets are highly
aliased and a reliable rotation period could not be determined, objects with a sufficiently
large number of observations would often yield convincing rotational light curves distinguishable
from nearby aliases.
Twenty-one of these high rotational amplitude objects were best fit with rotation periods 
greater than 24 hours, though in many of these cases additional aliases were 
possible. None of these lightcurves
has convincing detections of the cusp-like features and high amplitude drop outs 
suggestive of eclipsing by close or contact binaries.

(11351) Leucus remains an extreme outlier, with the longest well-determined period in our sample, and
one of the largest amplitude lightcurves. Our derived 
period of 445$\pm$10 hr is in good agreement
with the value from \citet{new_Mottola} of 445.683$\pm$0.007 hr. Only one other Trojan asteroid 
in our sample -- 65232 -- 
has a plausibly longer rotation period (480$\pm$10 hours), but the sampling of this asteroid is
limited and more data are required for a definitive determination.
As ZTF continues to operate it will become increasingly possible
to measure reliable periods and determine if (11351) Leucus truly is unique.

\section{Conclusion}\label{conclude}
This paper presents a new method
for extracting quantitative 
information from sparse photometric data
by using a probabilistic model of rotational
amplitude, allowing us to ignore the effects
of unknown rotational periods. We report color, phase parameter, absolute magnitude and amplitude of rotation measurements for 1049 Trojan asteroids
and examine the overall population statistics of our sample. 
We find that the phase parameter for Jupiter Trojan asteroids
is correlated with color, with average values of $0.14\pm 0.05$
and $0.22 \pm 0.07$ for the less-red and red populations, respectively.
The distribution of the amplitude of the rotational light curve appears relatively constant with asteroid size, and is not distinguishable from
the distribution for main-belt asteroids of similar sizes.

\acknowledgements
We have benefited from insightful discussion with James Keane, Anna Simpson, Samantha Trumbo,
and Elizabeth Bailey. We thank Stefano Mottola for sharing current photometry for comparison. 
We thank the two anonymous referees 
for a very thorough reading of this manuscript and
for multiple excellent suggestions for clarifying
the presentation.
This analysis based on observations obtained with the Samuel Oschin 48-inch Telescope at the Palomar Observatory as part of the Zwicky Transient Facility project. ZTF is supported by the National Science Foundation under Grant No. AST-1440341 and a collaboration including Caltech, IPAC, the Weizmann Institute for Science, the Oskar Klein Center at Stockholm University, the University of Maryland, the University of Washington, Deutsches Elektronen-Synchrotron and Humboldt University, Los Alamos National Laboratories, the TANGO Consortium of Taiwan, the University of Wisconsin at Milwaukee, and Lawrence Berkeley National Laboratories. Operations are conducted by COO, IPAC, and UW. 

\bibliographystyle{apj}
\bibliography{sources}

\end{document}